\documentclass[twocolumn,aps,showpacs]{revtex4} 
\usepackage{psfig}
\begin{document} 
\newcommand{\dprime}{{\prime\prime}}
\newcommand{\os}{{\overline\sigma}}
\newcommand{\be}{\begin{equation}}
\newcommand{\den}{\overline{n}} 
\newcommand{\ee}{\end{equation}}
\newcommand{\bea}{\begin{eqnarray}} 
\newcommand{\eea}{\end{eqnarray}}
\newcommand{\nn}{\nonumber} 
\newcommand{\vk}{{\bf k}}
\newcommand{\vE}{{\bf E}}
\newcommand{\vj}{{\bf j}}
\newcommand{\vs}{{\bf v}_s}
\newcommand{\vn}{{\bf v}_n}
\newcommand{\vv}{{\bf v}} 
\newcommand{\la}{\langle}
\newcommand{\ra}{\rangle} 
\newcommand{\ph}{\phi} 
\newcommand{\dg}{\dagger}
\renewcommand{\vr}{\bf{r}} 
\newcommand{\vq}{{\bf{q}}}
\newcommand{\vQ}{{\bf{Q}}} 
\newcommand{\hj}{\hat{\alpha}}
\newcommand{\hx}{\hat{\bf x}} 
\newcommand{\hy}{\hat{\bf y}}
\newcommand{\hz}{\hat{\bf z}}
\newcommand{\vS}{{\bf S}} 
\newcommand{\cV}{{\cal U}}
\newcommand{\cD}{{\cal D}} 
\newcommand{\tnh}{{\rm tanh}}
\newcommand{\sh}{{\rm sech}} 
\newcommand{\vR}{{\bf R}}
\newcommand{\crx}{c^\dg(\vr)c(\vr+\hx)}
\newcommand{\crkubox}{c^\dg(\vr)c(\vr+\hat{x})}
\newcommand{\pll}{\parallel} 
\newcommand{\crj}{c^\dg(\vr)c(\vr+\hj)}
\newcommand{\crmj}{c^\dg(\vr)c(\vr - \hj)}
\newcommand{\sumall}{\sum_{\vr}} 
\newcommand{\sumx}{\sum_{r_1}}
\newcommand{\nabj}{\nabla_\alpha \theta(\vr)} 
\newcommand{\nabx}{\nabla_1\theta(\vr)} 
\newcommand{\sumy}{\sum_{r_2,\ldots,r_d}}
\newcommand{\krj}{K(\vr,\vr+\hj)} 
\newcommand{\sigr}{|\psi_0\rangle}
\newcommand{\sigl}{\langle\psi_0 |}
\newcommand{\sier}{|\psi_{\Phi}\rangle}
\newcommand{\siel}{\langle\psi_{\Phi}|}
\newcommand{\sumrj}{\sum_{\vr,\alpha=1\ldots d}}
\newcommand{\krw}{K(\vr,\vr+\hx)} 
\newcommand{\Dtheta}{\Delta\theta}
\newcommand{\rhonew}{\hat{\rho}(\Phi)}
\newcommand{\rhoold}{\hat{\rho_0}(\Phi)} 
\newcommand{\dt}{\delta\tau}
\newcommand{\cP}{{\cal P}} 
\newcommand{\cS}{{\cal S}}
\newcommand{\vm}{{\bf m}} 
\newcommand{\hnr}{\hat{n}({\vr})}
\newcommand{\hnm}{\hat{n}({\vm})} 
\newcommand{\del}{\hat{\delta}}
\newcommand{\upa}{\uparrow} 
\newcommand{\dna}{\downarrow}
\title{
	Experimental implications of quantum phase fluctuations in layered 
	superconductors
}
\author {Arun Paramekanti
       }
\affiliation{
Dept. of Theoretical Physics, Tata Institute of Fundamental Research, Mumbai 400 005, India \\
         }

%\affiliation{
%\begin{minipage}[t]{6.0in}
\begin{abstract}
I study the effect of quantum and thermal phase fluctuations on the
in-plane and $c$-axis superfluid stiffness of layered $d$-wave
superconductors. First, I show that quantum phase fluctuations in the
superconductor can be damped in the presence of external screening
of Coulomb interactions,
and suggest an experiment to test the importance of these
fluctuations, by placing a metal in close proximity to the
superconductor to induce such screening. Second, I show that a
combination of quantum phase fluctuations and the linear temperature
dependence of the in-plane superfluid stiffness leads to a linear
temperature dependence of the c-axis penetration depth, below a
temperature scale determined by the magnitude of in-plane dissipation.
\typeout{polish abstract}
\end{abstract}
%\end{minipage}
%}

\maketitle

\section{Introduction}
%\noindent{\bf Introduction}:
The linear temperature dependence of the low temperature in-plane
penetration depth, $\lambda_{ab}(T)$ is well established now in nearly
all the high-$T_c$ superconductors \cite{hardy93,panagopoulos98}. This
behavior is most simply explained in terms of nodal quasiparticle
excitations in a d-wave superconductor \cite{nodal}.  These low energy
nodal quasiparticles then determine the low temperature behavior of
thermodynamic properties and in-plane response functions.  However,
since the high-$T_c$ superconductors (SC's) have a low superfluid density 
and a short coherence length compared to conventional superconductors, it 
is plausible that in addition to quasiparticle excitations, quantum and
thermal phase fluctuations could also be important.  The importance of
such quantum phase fluctuations, especially for $c$-axis properties, has
been emphasized earlier, from an analysis of the c-axis optical conductivity 
and sum rules\cite{ioffe99}.

Based on our study of an effective quantum XY phase action in the 
superconducting state, we had argued \cite{arun00,lara00} that quantum 
phase fluctuations are important in the cuprate SC's at low temperature.
With increasing temperature, in the presence of dissipation,
the system gradually crosses over from a quantum fluctuation regime to
a regime of classical phase fluctuations
\cite{lemberger00.theory,lara00} at a crossover scale, $T_{\rm cl}
\lesssim T_c$. (This has been observed in recent experiments on thin films of
conventional dirty $s$-wave superconductors\cite{lemberger00.expt}
where $T_{\rm cl} \sim 0.94~T_c$.) It has been suggested
\cite{carlson99}, that this crossover scale $T_{\rm
cl}\ll T_c$ in the high-$T_c$ systems, and thus classical phase fluctuations 
should dominate the
low temperature superconducting state properties.  However, for
parameter values relevant to the cuprate SC's, and even {\it overestimating} 
the magnitude of dissipation, which should lead to a {\em smaller} crossover
temperature, we found\cite{lara00} that $T_{\rm cl} \sim
20 K$. Since a linear $T$ behavior of the superfluid stiffness has been
observed in many high-$T_c$ SC's from $\sim 30 K$ down to $\sim 1 K$, 
with no evidence of such a thermal crossover, 
we conclude that the crossover temperature is high $\gtrsim
30 K$, and classical effects are irrelevant for in-plane 
properties at low temperatures $T \ll T_c$. The low temperature regime is 
thus described entirely by quasiparticle excitations around a ground state 
with strong quantum phase fluctuations. Recently, an experiment has been
proposed \cite{kwon00} to observe these quantum phase fluctuations
by studying the excess current along the $c$-axis, after suppressing
Josephson tunneling using an in-plane magnetic field.

The presence of strong quantum phase fluctuations in the low temperature
superconducting (SC) state raises several questions: Can the magnitude
of low temperature quantum phase fluctuations be directly probed?
What is the effect of such phase fluctuations on the $c$-axis
superfluid stiffness?  We shall study these questions in this
paper. The principal results of this paper are as follows. (1) External
screening of the Coulomb interactions in a superconductor damps
quantum phase fluctuations. Based on this I suggest an experiment where
a metal is placed in close proximity to the superconductor to induce
such external screening, similar to experiments which have been
carried on gated Josephson junction arrays\cite{rimberg97}.  In the
quantum fluctuation regime, relevant to our case, this screening reduces
the fluctuations and leads to observable changes in the superfluid
stiffness and its temperature dependence. 
In the classical regime, the screening would not lead to any
change in the superfluid stiffness, since the fluctuations are already
overdamped. Thus, this experiment would serve as a way to directly
measure the magnitude of quantum fluctuations in the
superconductor. (2) I show that quantum phase fluctuations can lead to
a linear $T$ increase of the $c$-axis penetration depth. This linear
$T$ slope arises from the linear $T$ dependence of the {\em in-plane
superfluid stiffness} and the Josephson coupling between layers. 
The magnitude of this effect is sensitive to the in-plane dissipation. 
The linear temperature dependence
may be consistent with some low temperature $c$-axis penetration
depth data \cite{shibauchi96,panagopoulos97,shovkun00} in
YBa$_2$Cu$_3$O$_{7-\delta}$ 
(YBCO) and Bi$_2$Sr$_2$CaCu$_2$O$_{8+\delta}$ (BSCCO) systems, 
thus serving as indirect
evidence of quantum phase fluctuations in the SC state. 
More generically however, the $c$-axis penetration depth does
not appear to have a linear temperature dependence. 
I discuss possible implications of this towards the end.

\section{Effective phase action}
%\noindent{\bf Effective phase action}:
I begin by briefly reviewing the formalism of
Refs~\cite{arun00,lara00} to study phase fluctuations in the low
temperature superconducting state.  We model the dynamics of the phase
variables $\theta_\vR(\tau)$ defined on a coarse-grained lattice (with
lattice spacing equal to the coherence length, $\xi_0$) by a quantum
XY action. We have derived the following coarse-grained action for
the layered d-wave SCs:
\bea
&&\!\!\!\!S[\theta]=\frac{1}{8 T}{\sum_{\vQ,\omega_n}}
\left[\frac{\omega^2_n \xi_0^2
d_c}{\tilde{V}(\vQ)}+\frac{\overline{\sigma}}{2\pi}
|\omega_n| \gamma_{_\pll}(\vQ)\right] |\theta(\vQ,\omega_n)|^2 \nn \\
\!\!\!\!&+&\!\!\!\!
\frac{1}{4}\!\int^{1/T}_0\!\!d\tau\sum_{\vR,\alpha}\!J_\alpha
\left\{ 1-\cos[\theta(\vR,\tau)-\theta(\vR+\alpha,
\tau)] \right\}
\label{xy1}
\eea
where $d_c$ is the interplane separation, $\xi_0$ is the in-plane
coherence length, $\alpha=x,y,z$ and 
$\gamma_{_\pll}(\vQ)=(4-2 \cos Q_x -2 \cos Q_y)$.
The internal dissipation in the superconductor arises from the 
conductivity of the electronic
degrees of freedom which have been integrated out, and is parametrized by 
$\overline{\sigma}$
where, for simplicity, I assume a constant (ohmic) layer conductance
$\sigma d_c=(e^2/h) \overline{\sigma}$.  The couplings $J_\alpha$ are
given by $J_{x,y}\equiv J_{_\pll}= D^0_{_\pll}d_c$ and $J_z\equiv
J_{_\perp}=D^0_{_\perp}d_c (\xi_0/d_c)^2$, where
$D^0_{_\pll},D^0_{_\perp}$ denote the bare in-plane and c-axis
stiffnesses respectively, which are related to the penetration depths
through $\lambda_{_\pll,_\perp}^{-2}=4\pi e^2 D_{_\pll,_\perp}/\hbar^2
c^2$.  $\tilde{V}(\vQ)\equiv V(\vQ_{_\pll} a/\xi_0,\vQ_{_\perp})$
denotes the scaled Coulomb interaction which arises from the coarse
graining procedure, with
\be
V(\vQ) = \frac{2 \pi e^2 a d_c}{Q_\pll\epsilon_b}
\left[\frac{\sinh(Q_\pll d_c/a)}{{\rm cosh}(Q_\pll d_c/a)-\cos Q_\perp
}\right]
\label{coulomb}
\ee
being the Coulomb interaction appropriate for layered systems. Here
$\epsilon_b$ is some background dielectric constant, $a$ is the
original in-plane lattice spacing, and $Q_\pll,Q_\perp$ are the in-plane and
c-axis components of the momentum, measured in units of $1/a$ and
$1/d_c$ respectively.

I analyze the quantum XY action within the self consistent harmonic
approximation \cite{wood82} (SCHA). The SCHA
replaces the above action by a trial harmonic theory with the
renormalized stiffness $D_{_\pll,_\perp}$ chosen to minimize the free
energy. It thus takes into account the
renormalization of the stiffness due to anharmonic longitudinal phase
fluctuations but ignores vortices and vortex-antivortex pairs 
(transverse fluctuations). The validity of this approximation at low
temperature, in the
classical case where only $n=0$ Matsubara frequency is retained,
has been confirmed in recent Monte Carlo simulations of the
classical XY model \cite{carlson99}. Carrying out the SCHA for our
anisotropic case leads to
\be
D_{_\pll,_\perp} = 
D^0_{_\pll,_\perp} \exp(- \la\delta\theta^2_{_\pll,_\perp}\ra/2)
\label{scha1}
\ee
where $\delta\theta_{_\pll,_\perp} \equiv
(\theta_{\vr,\tau}-\theta_{\vr+\alpha, \tau})$ with $\alpha=x,y$ for
the in-plane ($\pll$) case and $\alpha=z$ for the c-axis ($\perp$)
case. The expectation values are evaluated in the renormalized
harmonic theory, which yields
\bea
\la\delta\theta^2_{_\pll}\ra &=& 2 T \int_{-\pi}^{\pi}~\frac{d^3\vQ}{(2\pi)^3} 
\sum_{n}\gamma_{_\pll}(\vQ) G(\vQ,\omega_n) \nonumber \\
\la\delta\theta^2_{_\perp}\ra &=& 4 T \int_{-\pi}^{\pi}~\frac{d^3\vQ}
{(2\pi)^3} 
\sum_{n}\gamma_{_\perp}(\vQ) G(\vQ,\omega_n),
\label{scha2}
\eea
where $\gamma_{_\pll}(\vQ)=(4-2\cos Q_x-2\cos Q_y)$ and
$\gamma_{_\perp}(\vQ)=(2-2\cos Q_z)$.  Here, $G(\vQ,\omega_n)$ is the
phase propagator given by
\bea
G^{-1}(\vQ,\omega_n) 
= \frac{\omega^2_n\xi_0^2 d_c}{\tilde{V}(\vQ)}& +& \left(J_{_\pll} +
\frac{\overline{\sigma}}{2\pi} |\omega_n|\right) \gamma_{_\pll}(\vQ) \nn \\
&+& J_{_\perp} \gamma_{_\perp}(\vQ).
\label{scha3}
\eea 
In what follows, I analyze the above equations in various limits,
at low temperature, and also compare with a full self-consistent
numerical solution\cite{footnote.numerics} of Eqs.~(\ref{scha1}) -
(\ref{scha3}).

\section{An experimental probe of in-plane quantum phase fluctuations} 
In this Section, I study the damping of in-plane quantum phase
fluctuations due to {\it external} screening of Coulomb interactions.
This leads us naturally to a suggestion for an experiment to estimate the
magnitude of these fluctuations in the cuprate SC's.
Previous work on the damping of quantum phase fluctuations
include studies on resistively shunted Josephson junctions arrays
with short range charging energy
\cite{chakravarty86} and on a related model for superconductors with a 
low superfluid density \cite{emery95}.
These works focussed on the quantum phase transition from an 
insulator to a superconductor, driven by increasing the 
strength of dissipation.
Subsequently, such a quantum phase transition was observed experimentally 
in gated Josephson junction arrays \cite{rimberg97}, where the quantum 
phase fluctuations were damped by external screening from a (gate-tuned) 
metallic bath.
I show below within our phase fluctuation action, which differs in
important respects from the earlier models, that such damping of 
quantum fluctuations through
an external metallic bath leads to {\it observable consequences} for the 
low temperature superfluid stiffness. This may be used to study the 
importance of quantum fluctuations in the high-$T_c$ superconductors, 
with a set-up similar to the one used in the above experiment. 

Let us confine ourselves to the case of a thin superconducting film 
capacitively coupled to a metallic bath, and work in the two-dimensional 
($2D$) limit of the phase action. For electrons situated at an interface 
between vacuum, with a dielectric constant of unity, and a (metallic)
substrate with $\epsilon_{\rm sub}(\omega) = \left(\epsilon_\infty + 4\pi i 
\sigma_{\rm ext}/\omega \right)$, 
the Coulomb interaction is given by $V_{2D}(\vQ)=4 \pi 
e^2 a d_c/\left(1+\epsilon_{\rm sub}(\omega)\right) Q_{_\pll}$.
Thus, the metallic bath provides dynamical screening
of the Coulomb interactions\cite{emery95,depalo99}, and
leads to an additional term in
the phase action in Eq.~(\ref{xy1}), such that
\be 
(\os/2\pi)
\gamma_{_\pll} (\vQ)|\omega_n| \to \left[\os \gamma_{_\pll} (\vQ) +
\os_E \vQ_{_\pll}\right] \frac{|\omega_n|}{2\pi}, 
\ee 
where $\os_E \equiv (\xi_0/d_c)~(\sigma_{\rm ext} d_c)/(e^2/h)$, and
$\epsilon_b \to (1+\epsilon_\infty)/2$ in the Coulomb interaction
in Eq.~(\ref{coulomb}). Thus, external dissipation 
due the metallic bath ($\sigma_{\rm ext}$) appears together with the
internal dissipation from the electronic degrees of freedom of the
superconductor which have been integrated out ($\os$), 
and they can both lead to damping of quantum fluctuations in a similar
manner. Note that our approach differs from the earlier work of
Ref.~\cite{emery95} in two important ways.
(i) We retain the dynamical term $\omega^2_n \xi^2_0 d_c/\tilde{V}(\vQ)$ 
in the action in Eq.~(\ref{xy1}), and the renormalized propagator in
Eq.~(\ref{scha3}). (ii) We make a distinction between the dissipation 
which arises from the degrees of freedom internal to the superconductor 
and that from external screening. This is reflected in the fact
that the conductivities $\os$ and $\os_E$ appear with different 
$\vQ$-dependent coefficients in the above action \cite{footnote.gauge}. 

\begin{figure}
\begin{center}
\vskip-2mm
\hspace*{0mm}
\psfig{file=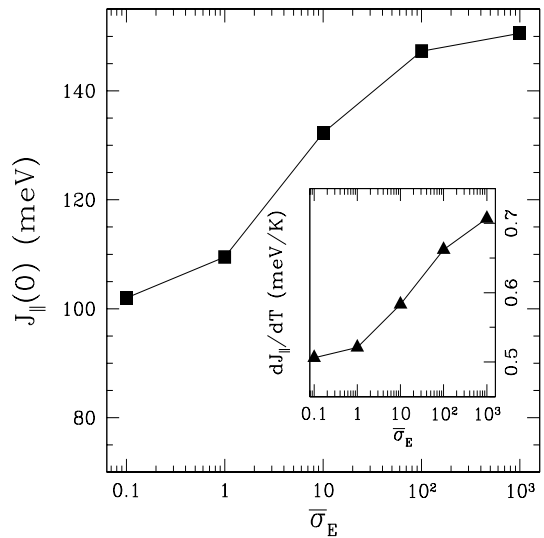,height=2.8in,width=2.8in,angle=0}
\vskip1mm
\caption{The $T=0$ in-plane superfluid stiffness for 
various values of external dissipation $\overline{\sigma}_E$. The inset
shows the behavior of the linear $T$ slope of $J_{_\pll}(T)$.
The above results have been obtained by solving the
Eqs.~(\ref{scha1})-(\ref{scha3}) for a two dimensional case, assuming
bare bilayer stiffness $J_{_\pll}(0) \approx 150 meV$ with a slope 
$\sim 0.65 meV/K$, and an internal
dissipation $\os=10$. These values have been chosen such that, for 
$\os_E=0$ we obtain a penetration
depth $\lambda_{_\pll}(0) \approx 1600 \AA$, and a slope
$d\lambda_{_\pll}/dT \approx 4 \AA/K$, in agreement with 
experiment\protect\cite{hardy93}.}
\label{fig1}
\end{center}
\end{figure}

To quantitatively study the effect of the external screening, let
us consider parameter values relevant to the cuprate SC's. I choose
$\epsilon_\infty \approx 10$, $\xi_0/a\approx 10$ and $\os \approx 10$
as representative of the YBCO system \cite{footnote} and any typical
substrate material. We model
YBCO as a system of strongly coupled bilayers, with the bilayer
stiffness being twice the single layer stiffness, and an inter-bilayer
spacing $d_c/a \approx 3$ being twice the mean layer spacing. This will
simplify our calculation, since we do not have to introduce an
additional parameter to distinguish intra- and inter-bilayer couplings;
a more sophisticated calculation would not lead to any qualitative or 
significant quantitative change. For the present analysis,
we will use a bare {\it bilayer} stiffness and its linear $T$ slope
such that the renormalized stiffness $J_{_\pll}(T)$, for the above
parameters and with $\os_E =0$, leads to a penetration depth
$\lambda_{_\pll}(0) \approx 1600 \AA$ and $d\lambda_{_\pll}/dT \approx
4 \AA/K$ in agreement with experiment\cite{hardy93,hardy98}.  We can then
vary the external dissipation $\os_E$, and study its effect on
$J_{_\pll}(T)$.  In Fig.~\ref{fig1}, I plot the
behavior of the the stiffness $J_{_\pll}(0)$, and its
slope $d J_{_\pll}/d T$, for various values of $\os_E$
corresponding to differing levels of external dissipation. 

It is clear that the stiffness $J_{_\pll}(0)$ increases with increasing
dissipation, as quantum fluctuations are damped, leading to a more
ordered state. With increasing dissipation, (classical) thermal phase
fluctuations also contribute to the normal fluid density, in addition 
to the 'bare' quasiparticle contribution, leading to an 
enhancement in the slope of $J_{_\pll}(T)$. A measurement of 
the stiffness and its temperature dependence in the presence
of external screening would thus serve as a test of quantum 
phase fluctuations. This could be probed in low frequency optical
experiments, or penetration depth experiments. A minor caveat concerning
penetration depth experiments is that the effective 
(Pearl) penetration depth $\lambda_{\rm eff} = \lambda^2_{_\pll}/\delta$ 
where $\delta$ is the film thickness, and hence one has cannot use the
strictly $2D$ limit in the above calculation but $\delta$ has to be kept
finite. This is not expected to lead to any qualitative changes in
our results for $J_{_\pll}(T)$.

\section{Quantum phase fluctuations and the c-axis penetration depth} 
%\noindent {\bf $c$-axis penetration depth}:
Having shown that quantum phase fluctuations may affect the in-plane
superfluid stiffness, I next turn to examine the effect of these
fluctuations on the $c$-axis superfluid stiffness.  Many earlier
studies of the $c$-axis stiffness have focussed on the quasiparticle
contribution to the normal fluid density, emphasizing the role of
tunneling matrix elements\cite{xiang96}, or effects of disorder and
pair tunneling on the temperature dependence of the critical
current\cite{klemm95,levin96}. Here, I study the effect of phase
fluctuations on the c-axis superfluid stiffness within the SCHA, and
find that phase fluctuations can lead to a linear temperature
dependence of the $c$-axis superfluid stiffness and penetration
depth. This linear $T$ slope arises from the linear $T$ dependence of
the {\it in-plane stiffness} and the Josephson coupling between
layers, and it is a {\it generic} effect in such models.

We shall begin by assuming that the bare c-axis stiffness is
$T$-independent at low temperature. This assumption may be justified
as follows.  Quasiparticles may be important for in-plane properties,
but the inter-plane tunneling matrix element\cite{caxis} being
proportional to $\sim (\cos k_x - \cos k_y )^2$, leads to a very small
depletion of the superfluid density, since matrix element vanishes for
in-plane nodal quasiparticles.  In our calculation we shall neglect
$c$-axis dissipation since the sharp Josephson plasmon seen in
experiments on BSCCO implies a very small $c$-axis conductivity, and
the conductivity has been measured to be small over a wide frequency
range \cite{uchida92}. In YBCO, there are added complications due
to the chains, but I will ignore this.

To study the effect of phase fluctuations, let us analyze the
fluctuation integral for $\la\delta \theta^2_{_\perp}\ra$ in
Eq.~(\ref{scha2}) for the physically relevant case of
$J_{_\perp}/J_{_\pll}\ll 1$.  Since we are interested only in
$J_{_\perp}(T)$ we fix the in-plane stiffness, $J_{_\pll}(T)$, from
experiment, and study the $c$-axis fluctuations
$\la\delta\theta^2_{_\perp}\ra (T)$.  I will first present results for the
case with no in-plane dissipation ($\os = 0$) and later
consider the effect of dissipation to see how these results are
affected by a finite $\os \neq 0$. 
Finally, I will compare our results with experimental data
on BSCCO and YBCO.

\subsection{Dissipationless case ($\os=0$)}
For $\os=0$, we can first do the Matsubara summation in the fluctuation
integral in Eq.~(\ref{scha2}). This leads to 
\bea
\la\delta \theta^2_{_\perp}\ra&=&2 \int_{-\pi}^{\pi} \frac{d^3\vQ}
{(2\pi)^3} \gamma_{_\perp}\sqrt{\frac{\tilde{V}}{\left(
J_{_\pll} \gamma_{_\pll} +
J_{_\perp}\gamma_{_\perp}\right)}} \nonumber \\
&\times&\coth \left( \frac{1}{2 T}
\sqrt{\tilde{V}\left( J_{_\pll} \gamma_{_\pll}+
J_{_\perp}\gamma_{_\perp}\right)} \right),
\eea
where we have suppressed the $\vQ$ dependence of
$\gamma_{_\perp,_\pll}$ and $\tilde{V}$.  We next note that the
temperature dependence of $\la\delta \theta^2_{_\perp} \ra$ arises
from two sources: (i) in the $\coth$ factor, which corresponds to
thermal excitations of the plasmon mode, and (ii) in the prefactor,
through the temperature dependence of $J_{_\pll}$. I have numerically
checked that the $\coth$ factor may be set to unity at low temperature
for the cases of interest, since the in-plane plasmon energy is very
large. Even when the $c$-axis Josephson plasmon energy is very low, as
in case of BSCCO, the phase space for this low energy excitation is
small in the fluctuation integral and the characteristic plasmon
energy is very high (see Fig.~1 of Ref.\cite{arun00}).  The crossover
associated with this factor then only leads to large power laws for
temperatures above the $c$-axis plasmon scale. Further,
$J_{_\perp}/J_{_\pll}\ll 1$ means we can now safely set $J_{_\perp}=0$
in the prefactor. Thus we are led to
\bea 
\la\delta
\theta^2_{_\perp}\ra & \approx & \frac{2}{\sqrt{J_{_\pll}(T)}}
\int_{-\pi}^{\pi}\frac{d^3\vQ}{(2\pi)^3}
\gamma_{_\perp}(\vQ)\sqrt{\frac{\tilde{V}_\vQ} {\gamma_{_\pll} (\vQ)}}
\nonumber \\ 
& \equiv & C_1 \left[\left(\frac{2\pi
e^2}{\epsilon_b\xi_0}\right) \frac{1}{J_{_\pll}(T)}\right]^{1/2}
\eea 
where $C_1$ is a constant of order unity, which depends only on
$\xi_0/d_c$, and which can be determined numerically for a given
system.  It is now easy to see that the linear $T$ dependence of
$J_{_\pll}$ directly leads to a linear $T$ dependence of
$\la\delta\theta^2_{_\perp}\ra$ and hence of
$J_{_\perp},\lambda_{_\perp}$. To relate the slope of
$\lambda_{_\perp}$ to the slope of $J_{_\pll}$, we set $J_{_\pll}(T) =
J_{_\pll}(0) - \alpha T$. Using Eq.(\ref{scha1}) we then get
\be
\frac{d\lambda_{_\perp}}{d T} = C_1
\left(\frac{\lambda_{_\perp}(0)}{8} \right)
\left(\frac{\alpha}{J_{_\pll}(0)}\right) \sqrt{\left( \frac{2\pi
e^2}{\epsilon_b\xi_0} \right) \frac{1}{J_{_\pll}(0)}} 
\ee 

\subsection{Non-zero dissipation ($\os\neq 0$)}
%\noindent{\bf Non-zero dissipation, $\os \neq 0$}:
In the presence of dissipation, we can again analyze the Matsubara
summation in the fluctuation integral to obtain the asymptotic low
temperature behavior, similar to our earlier analysis for the in-plane
fluctuations \cite{lara00}. In this case, for $\os \gg 1$ and $T\to
0$, we find
\bea
\la\delta\theta^2_{_\perp}\ra&\approx&\frac{8}{\os} \ln \left(
\frac{\os}{2\pi} \sqrt{\frac{2\pi e^2} {\epsilon_b}
\frac{1}{J_{_\pll}(T)}}\right) \nonumber \\
& + & T^2 \frac{2 \os}{3} \int_{-\pi}^{\pi}
\frac{d^3\vQ}{(2\pi)^3}\frac{\gamma_{_\perp}
\gamma_{_\pll}}{(\gamma_{_\pll} J_{_\pll}+\gamma_{_\perp}
J_{_\perp})^2}
\eea

While the leading linear $T$ dependence of
$\la\delta\theta^2_{_\perp}\ra$ arises from the linear $T$ dependence
of $J_{_\pll}$ in the first term in the above equation, a $T^2$
dependence arises, primarily from the second term for $\os \gg 1$. 
With $J_{_\pll}(T)=J_{_\pll}(0)-\alpha T$, as before, the leading
temperature dependence of the fluctuations is now given by
\be
\la\delta\theta^2_{_\perp}\ra(T) - \la\delta\theta^2_{_\perp}\ra(0) =
\frac{4 T}{\os}\frac{\alpha}{J_{_\pll}(0)} + C_2 \frac{2\os}{3}\left(
\frac{T}{J_{_\pll}(0)}\right)^2 
\ee 
where 
\be 
C_2 =
\int_{-\pi}^{\pi}\frac{d^3\vQ}{(2\pi)^3}
\frac{\gamma_{_\pll}\gamma_{_\perp}} {\left(\gamma_{_\pll} +
(J_{_\perp}(0)/J_{_\pll}(0)) \gamma_{_\perp}\right)^2} 
\ee 
is a constant depending on the anisotropy $J_{_\perp}/J_{_\pll}$ at
$T=0$, which may be easily determined numerically.  This leads to a
crossover scale $T_2=4 \alpha J_{_\pll}(0)/(C_2\os^2)$ beyond which
temperature the linear temperature dependence crosses over to a $T^2$
behavior. For $T \ll T_2$, we find a linear $T$ behavior in
$\lambda_{_\perp}(T)$ with a slope 
\be
\frac{d \lambda_{_\perp}}{d T} = \frac{\alpha}{\os J_{_\pll}(0)}
\lambda_{_\perp}(0).
\ee
However, since the linear $T$ to quadratic $T^2$ crossover temperature
depends very sensitively on the dissipation, large dissipation
might lead to a $T^2$ behavior down to the lowest observed
temperatures. 
%In this connection, we note that varying the doping
%might affect the dissipation which could also result in a change in
%the temperature dependence of $\lambda_{_\perp}$ from $T$ to $T^2$.
We therefore make estimates of this temperature scale $T_2$ for BSCCO and
YBCO$_{6.95}$; we then compare our results for the temperature dependence
of $\lambda_{_\perp}(T)$ with some experimental data on BSCCO and
YBCO$_{6.95}$.

\subsection{Comparison with experiments}
%I have shown that an observation of linear-$T$ behavior in the $c$-axis 
%penetration depth would serve as an indirect measure of quantum phase
%fluctuations in the low temperature SC state. 
%Our next goal is to estimate these linear-$T$ slopes in BSCCO and YBCO 
%materials. 
I begin by fixing the parameters connected to the in-plane stiffness
in the phase action, and $\lambda_{_\perp}(0)$ since we are only 
interested in our predictions for the temperature dependence of 
$\lambda_{_\perp}(T)$. I then compare the slope
$d\lambda_{_\perp}(T)/dT$ with existing experiments in BSCCO and YBCO 
systems.

For BSCCO, I use $\xi_0/a \approx 10$ and $d_c/a \approx 4$, and
set the bilayer stiffness $J_{_\pll}(0)\approx 80 meV$ with its linear
$T$ slope $\alpha \approx 0.8 meV/K$. These values correspond
\cite{sflee96} to a penetration depth $\lambda_{_\pll}=2100 \AA$ 
with $d\lambda_{_\pll}/dT \approx 10 \AA/K$. 
I then set $\lambda_{_\perp}(0) \approx 150 \mu m$ and choose a 
reasonable value ($\os = 20$) for the internal dissipation. 
This gives $C_1 \approx 0.6$, $C_2 \approx 1.35$, and $T_2 \approx 60 K$ 
below which we expect to see linear $T$ behavior; the low 
temperature slope $d\lambda_{_\perp}/dT \approx 0.075 \mu m/K$ is 
somewhat smaller than some experimentally reported values 
\cite{shibauchi96,shovkun00} of $\sim 0.25$-$0.3 \mu m/K$. 

For YBCO$_{6.95}$, I use $\xi_0/a\approx 10$, $d_c/a\approx 3.2$,
and set the bilayer stiffness $J_{_\pll}(0)\approx 100
meV$ with its slope $\alpha \approx 0.5 meV/K$. This leads to 
a penetration depth $\lambda_{_\pll}\approx 1600 \AA$ with
$d\lambda_{_\pll}/dT \approx 4 \AA/K$ in agreement with
experiment\cite{hardy93,hardy98}. Finally, I set $\lambda_{_\perp} (0) 
\approx 1.1 \mu m$, and use a reasonable value $\os=10$ for the
intrinsic dissipation.
This gives $C_1 \approx 0.5$, $C_2 \approx 0.4$, and $T_2 \gtrsim T_c$, 
below which we expect to see linear $T$ behavior. Thus, the linear $T$ 
behavior from phase fluctuations is expected to persist over a larger
temperature scale in YBCO$_{6.95}$. I then find the slope
$d\lambda_{_\perp}/dT \approx 5 \AA/K$. This is somewhat smaller than 
some reports \cite{panagopoulos97} of $d\lambda_{_\perp}/dT \sim 
15$-$20 \AA/K$ on this
system. While this behavior was attributed \cite{panagopoulos97} to the 
effect of the chains in YBCO, the role of phase fluctuations could 
clearly also be important. The fluctuations at $T=0$
lead to a $\sim 30\%$ renormalization of the stiffness $J_{_\perp}(0)$, 
in reasonable
agreement with an earlier $c$-axis conductivity sum rule analysis
\cite{ioffe99} carried out for YBCO$_{6.6}$.

More generically, the experimentally observed $c$-axis penetration depth 
in BSCCO and YBCO$_{6.95}$ is reported to have a weaker 
temperature dependence (see Ref.~\cite{hardy98}
for instance), possibly $\sim T^2$ at low temperature. One possible reason
for this discrepancy between the prediction of the quantum phase
fluctuation model and the experiments could be the effects of disorder 
between the planes which deserves a more careful investigation.

%\begin{figure}[t]
%\begin{center}
%\includegraphics[width=7cm, angle=0]{fig2.eps}
%\caption{\footnotesize Comparison of the experimental $c$-axis
%penetration depth data\protect\cite{hardy98} (open squares) with the
%calculated phase fluctuation renormalized penetration depth (solid line).
%The solid line is obtained by solving the SCHA equations (\ref{scha1})-
%(\ref{scha3}), for parameter values relevant to bilayer YBCO$_{6.95}$ 
%and a dissipation parameter $\os \approx 10$. See text for details and
%discussion.}
%\label{fig2}
%\end{center}
%\end{figure}
\bigskip
\section{conclusions}
We have seen that it is possible to directly probe the importance
of in-plane quantum phase fluctuations in the high-$T_c$ superconductors
for $T \ll T_c$ through measurements of the in-plane penetration depth
$\lambda_{_\pll}(T)$ in the presence of external screening. I have also 
shown that an indirect measure of
quantum phase fluctuations may be obtained from studying the
temperature dependence of the $c$-axis penetration depth,
$\lambda_{_\perp}(T)$. It is possible that disorder between the planes 
could affect our prediction, leading to a weaker temperature dependence
more consistent with the experimental data; we leave this issue for
future work.
However the linear $T$ behavior might still be observable in some clean 
materials. Experimental tests of these would lead to a better understanding 
of the superconducting ground state and its low energy excitations.

\bigskip

{\noindent \bf Acknowledgements:} 
I am grateful to M. Randeria for his constant encouragement, useful
discussions, and critical comments on the paper. I thank D.
Gaitonde and C. Panagopoulos for helpful discussions.


\begin{thebibliography}{999}
\bibitem{footnote.numerics}
In the numerics, the Matsubara sum in Eq.~(\ref{scha2}) is cutoff at
finite $n$. While this affects the magnitude of quantum fluctuations
at $T=0$ to a small extent, the results for the temperature dependence
are not sensitive to this cutoff.
\bibitem{footnote.gauge}
This difference arises from the fact that the
external (bath) electrons have been treated in a gauge invariant
fashion, while gauge invariance is not recovered for the superconductor 
unless the phase fluctuations are also integrated out. 
See Ref.~\cite{arun00} for a detailed discussion of this last point.
This distinction between the two sources of dissipation was missed in 
earlier work \cite{emery95}.
\end{thebibliography}
\end{document}